\documentclass[11pt,twoside]{article}
\usepackage{./asp2014}

\aspSuppressVolSlug
\resetcounters

\begin{document}

\title{Solar ALMA Observations -- A new view of our host star}

\author{Sven Wedemeyer,$^{1,4}$ Tim Bastian,$^2$ Roman Braj\v sa,$^3,4$ Miroslav Barta,$^4$ Masumi Shimojo$^5$, Antonio Hales,$^6$ Pavel Yagoubov,$^7$ Hugh Hudson,$^{8,9}$ 
\affil{$^1$Institute of Theoretical Astrophysics, University of Oslo, Norway;  \email{sven.wedemeyer@astro.uio.no}}
\affil{$^2$National Radio Astronomy Observatory (NRAO), Charlottesville, VA, USA} 
\affil{$^3$Hvar Observatory, Faculty of Geodesy, University of Zagreb, Croatia}
\affil{$^4$European ALMA Regional Center - Czech node, Astronomical Institute of Academy of Sciences, Ondrejov, Czech Republic}
\affil{$^5$National Astronomical Observatory of Japan (NAOJ),  Japan} 
\affil{$^6$Joint ALMA Observatory (JAO), Santiago, Chile} 
\affil{$^7$European Organisation for Astronomical Research in the Southern Hemisphere (ESO), Garching bei M\"unchen,  Germany} 
\affil{$^8$UC Berkeley, USA} 
\affil{$^9$University of Glasgow, UK} 
}

% This section is for ADS Processing.  There must be one line per author.
\paperauthor{Sven~Wedemeyer}{sven.wedemeyer@astro.uio.no}{http://orcid.org/0000-0002-5006-7540}{University of Oslo}{Institute of Theoretical Astrophysics}{Oslo}{}{0315}{Norway}
%\paperauthor{Sample~Author2}{Author2Email@email.edu}{ORCID_Or_Blank}{Author2 Institution}{Author2 Department}{City}{State/Province}{Postal Code}{Country}
%\paperauthor{Sample~Author3}{Author3Email@email.edu}{ORCID_Or_Blank}{Author3 Institution}{Author3 Department}{City}{State/Province}{Postal Code}{Country}

\begin{abstract}
ALMA provides the necessary spatial, temporal and spectral resolution to explore central questions in contemporary solar physics with potentially far-reaching implications for stellar atmospheres and plasma physics.
It can uniquely constraint the thermal and magnetic field structure in the solar chromosphere with measurements that are highly complementary to simultaneous observations with other ground-based and space-borne instruments. 
Here, we highlight selected science cases. 
\end{abstract}

%================================================================================
%================================================================================
%================================================================================
\section{ALMA observations of the solar chromosphere} 
%================================================================================
%================================================================================
%%================================================================================
%
The solar radiation emitted at ALMA wavelengths originates mostly from the chromosphere -- a complex and dynamic (interface) region embedded between the photosphere and the corona. 
This region plays a key role in the transport of energy and matter and thus in the heating of the Sun's outer layers. 
Despite decades of intensive research, however, our understanding of the chromosphere is still elusive, which can be attributed to the rather small number of currently available suitable diagnostics and the challenges with their meaningful interpretation. 
ALMA now serves as a new tool with many impressive capabilities, promising significant progress in solar physics. 
Next to the high spatial, temporal, and spectral resolution, the following properties of ALMA yield large diagnostic potential for solar observations: 
(i)~ALMA measurements of the solar chromosphere serve as a linear thermometer, enabling the determination of the thermal structure of the chromosphere. 
(ii)~The formation height of the continuum radiation increases with wavelength, facilitating observations in different atmospheric layers. 
(iii)~Radio recombination lines (e.g., H and O\,VI) and molecular lines (e.g., CO) provide complementary kinetic and thermal diagnostics. 
(iv)~The polarisation of the continuum intensity and the Zeeman effect can be exploited for valuable chromospheric magnetic field measurements. 

%%================================================================================
\begin{figure}
\centering
\resizebox{11.6cm}{!}{\includegraphics[]{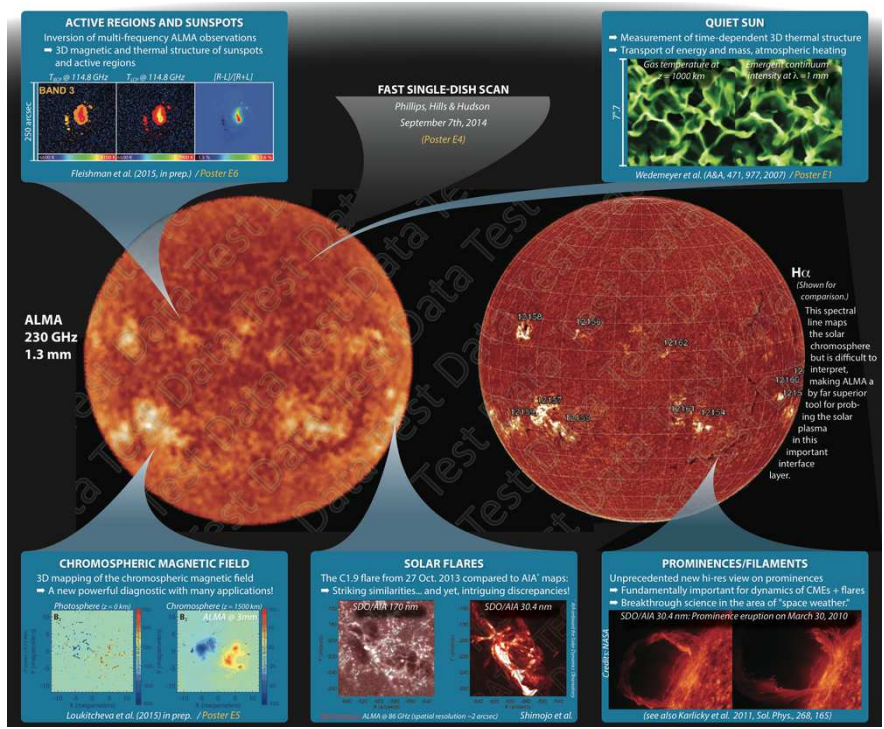}}
    \vspace*{-2mm}
    \caption{The large map in the middle left is the result of a scan over the full solar disk with a 
    single ALMA total power antenna at a frequency of 230\,GHz \citep{Phillips2015}. 
    Despite the limited resolution, the map shows many features which are visible in H$\alpha$, too (right middle). 
    The small insets demonstrate potential science cases based on numerical simulations and observations at other wavelengths.
        \label{fig1}
}
%\end{center}
\end{figure}
%%================================================================================

These impressive capabilities open up a large range of topics in solar physics 
\citep[see Fig.~1 and the upcoming review by the SSALMON group,][]{2015arXiv150205601W,ssalmon_ssrv15}. 
Among the central questions for ALMA are: 
(i)~Coronal and chromospheric heating -- measuring the 3D thermal structure and dynamics of the solar
chromosphere and thus sources and sinks of atmospheric heating; 
(ii)~Solar flares -- probing non-thermal radiation due to acceleration of high-energy electrons and secondary relativistic positrons; 
(iii)~Space weather -- formation, destabilisation and eruption of coronal filaments and prominences.  
Significant progress in these areas can be expected from ALMA.  

%================================================================================
%================================================================================
%================================================================================

\end{document}